\documentclass{elsart}

\advance\hoffset by -0.5in
\advance\textwidth by 1in
\advance\voffset by -0.4in
\advance\textheight by 0.8in
\usepackage{graphicx}

\begin{document}
\begin{frontmatter}

\title{Marvel Universe looks almost like a real social network}

\thanks{This work has been partly supported by the DGES, grant BFM2000-1113-C02-01.}

\author{R. Alberich, J. Miro-Julia, F. Rossell\'o}

\address{Departament de Matem\`atiques i Inform\`atica,
Universitat de les Illes Balears,\\
07071 Palma de Mallorca (Spain)\\
\emph{\texttt{$\mathtt{\{}$ricardo,joe,cesc$\mathtt{\}}$@ipc4.uib.es}}}

\begin{abstract}
We investigate the structure of the Marvel Universe collaboration 
network, where two Marvel characters are considered linked if they 
jointly appear in the same Marvel comic book.  We show that this 
network is clearly not a random network, and that it has most, but not 
all, characteristics of ``real-life'' collaboration networks, such as 
movie actors or scientific collaboration networks.  The study of this 
artificial universe that tries to look like a real one, helps to
understand that there are underlying principles that make real-life 
networks have definite characteristics.
\end{abstract}

\end{frontmatter}

\section{Introduction}

A recent popular topic of research in social networks has been the 
study of collaboration networks.  In these, the vertexes (or nodes) 
represent people and the edges that link pairs of nodes denote the 
existence of some kind of collaboration between them.  Their 
popularity stems mainly from two factors.  First, they are more 
objective than other social networks like friendship or 
first-name-knowledge networks.  Their links have a definite meaning, 
while, for instance, the meaning of links in friendship networks is 
subjective and thus possibly non-homogeneous throughout.  And second, 
the existence and availability of large databases containing all 
information concerning movies, baseball teams, scientific papers, and 
other large fields of collaboration, makes it easier to create and 
study these networks, while reliable friendship networks can only be 
raised through the intensive gathering of information by means of 
interviews.  Furthermore, the databases from which collaboration 
networks are extracted usually contain information about the time when 
each collaboration has taken place.  This information can be used to 
describe the evolution of the network and then to extract properties 
about how social networks grow \cite{BJNRSV,N4}.

A well-known collaboration network is the \emph{Movie Actors network}, 
also dubbed the \emph{Hollywood network}.  In it, nodes represent 
actors and actresses, and a link is added between two nodes when they 
have jointly appeared in the same film.  All information concerning 
this network is accessible at the \emph{Internet Movie Database} 
\cite{IMDB}, and it has been studied from a mathematical point of 
view \cite{ASBS-00,NWS,Watts,W-S}.  This is the basis of the popular 
\emph{Kevin Bacon game} \cite{KBgame}, which consists of trying to 
connect any given actor or actress to Kevin Bacon through the shortest 
possible path of collaborations in films.

\emph{Scientific collaboration networks} have also been thoroughly 
studied in the last years.  In such a network, nodes represent 
scientists and links denote the coauthorship of a scientific piece of 
work contained in some database.  For instance, there is the so-called 
{Erd\"os collaboration graph}.  Paul Erd\"os was a Hungarian 
mathematician, dead in 1996, who published over 1500 papers with 492 
coauthors, more than any other mathematician in history.  The 
\emph{Erd\"os collaboration graph} is the mathematicians' 
collaboration network around Erd\"os himself \cite{BM00,CG99}, built 
up through data collected by Grossman \cite{ErdGr}.  Also, Newman \cite{N1,N2,N3,N4} 
has studied in detail the scientific collaboration networks 
corresponding to several databases, namely MEDLINE (biomedical 
research papers in refereed journals), SPIRES (preprints and published 
papers in high-energy physics), NCSTRL (preprints in computer 
science), and Los Alamos e-Print Archive (preprints in physics).  
Bar\'abasi et al.\ \cite{BJNRSV} have studied the networks based on two databases 
containing articles on mathematics and neuro-science, respectively, 
published in relevant journals.

Newman \cite{N1} argues that scientific collaboration networks are true 
social networks, since most pairs of scientists that have written a 
paper together are genuinely acquainted with one another.  The social 
meaning of the Hollywood network is, in this sense, weaker, because it 
has been built up mainly through the decisions of cast directors, 
producers and agents, rather than the voluntary collaboration of 
actors.  Despite these, and other, differences, 
all collaboration networks studied so far present the same basic 
features: (a) on average, every pair of nodes can be connected through 
a short path within the network; (b) the probability that two nodes 
are linked is greater if they share a neighbor; and (c) the fraction 
of nodes with $k$ neighbors decays roughly as a function of the form 
$k^{-\tau}$ for some positive exponent $\tau$, with perhaps a cutoff 
for large values of $k$.  A network satisfying properties (a) and (b) 
is called a \textsl{small world} \cite{Watts,W-S}, and if it 
satisfies (c) then it is called \emph{scale-free} 
\cite{ASBS-00,Barab-sf}.

Does this similarity in features represent some profound principle in 
human interaction?  Or, on the contrary, does any large network with 
some ``collaboration'' between nodes present these characteristics?  A 
first, theoretical, step in this direction has been recently made by 
Newman et al.\ \cite{NWS}, who have developed a theory of random collaboration 
networks and have shown that some statistical data of most 
``real-life'' collaboration networks differ substantially from the 
data obtained from random models.

In this paper we want to contribute to a possible answer to these 
questions by analyzing a new collaboration network, that is 
artificial, but mimics real-life networks: the \emph{Marvel Universe} 
collaboration network.  In it, the nodes correspond to Marvel Comics 
characters, and two nodes are linked when the corresponding characters 
have jointly appeared in the same Marvel comic book.

Marvel Comics, together with DC Comics, have been for many decades the 
two main comic book publishing companies in the world 
\cite{Daniels,Lavin}.  It was founded in 1939 by M. Goodman, under 
the name of Timely Comics Inc.; it changed its name in the early 1960s 
to Marvel Comics, which was also the name of the first comic book 
published by Timely.  After a first decade of popularity, known as the 
Golden Age of comics (1939-49), and a later period of general waning 
of interest in super-hero stories, Marvel relaunched in 1961 its 
super-hero comic books publishing line, starting what has been known 
as the \emph{Marvel Age of Comics}.  Some of the characters created in 
this period, like Spider-Man, the Fantastic Four, the X-Men, together 
with other characters rescued from the Golden Age, like Captain 
America, are world-wide known and have become cultural icons of the 
western society of the last forty years.

One of the main features of Marvel Comics from the sixties to our days 
has been the creation and development, under the leading pen of Stan 
Lee, of the so-called {Marvel Universe}.  Although \emph{crossovers} 
(a hero with its own title series appears in an issue of another 
hero's series) were not uncommon in the Golden Age period, the nature 
and span of the crossovers in the books from the Marvel Age led to the 
perception that all Marvel characters lived their adventures in the 
same fictional cosmos, called the \emph{Marvel Universe}, where they 
interacted like real actors.  This concept was helped by the 
interrelation of all titles that were being created, which made 
characters and even plots cross over on a regular basis, by the 
appearance of the same villains and secondary characters in comic 
books of different titles, and by continuous references to events that 
were simultaneously happening, or had happened, in other books.  A 
paradigm of the Marvel Universe could be Quicksilver, who appeared 
first as a member of Magneto's Brotherhood of Evil Mutants in the 
early issues of {Uncanny X-Men}, then he became a member of the 
Avengers and later of X-Factor, to end as the leader of the Knights of 
Wundagore; he is also the son of Magneto, the twin brother of the 
Scarlet Witch, and he married Crystal, a former fianc\'ee of Fantastic 
Four's Human Torch and a member of the Inhumans (as well as of the 
Fantastic Four as a substitute of the Invisible Woman when she took 
her ``maternal leave'').

The Marvel Universe network captures the social structure of this 
Marvel Universe, because most pairs of characters that have jointly 
appeared in the same comic book have fought shoulder to shoulder or 
each other, or have had some other strong relationship, like family 
ties or kidnapping.  Thus, it shares, in its artificial way, the true 
social nature of scientific collaboration networks, while the way it 
has grown has echoes of the Hollywood network, as writers, directors 
and producers create their characters and assign them to actors in a 
way that somewhat resembles the way Marvel writers make characters  
appear in comic books.

Thus, besides any sentimental or cultural motive, this is where the 
main reason for studying the properties of the Marvel Universe lies: 
it is a purely artificial social network, whose nodes correspond to 
invented entities and whose links have been raised by a team of 
writers without any preconception for a period of forty years.  We 
considered therefore interesting to know if the Marvel Universe 
network's artificial nature would resemble real-life collaboration 
networks, or, on the contrary, would rather look like a random 
collaboration network.  As we shall see, the first is essentially the 
case: most statistical data of the Marvel Universe differ from the 
random model in a way reminiscent of real-life collaboration networks.  
Nevertheless, we must mention that there is one particular value, the 
clustering coefficient, that also greatly differs from what one would 
expect in a real-life collaboration network.  We shall argue that this 
difference stems from the way how characters were distributed among 
books by Marvel writers, which is different from the way how real-life 
scientists join to write scientific papers.  After all, men, 
even Stan Lee (The Man himself) cannot imitate society.

\section{The Marvel Universe  network}
We define the \emph{Marvel Universe network} (\emph{MU}) as the network whose 
nodes are significant Marvel characters and where two characters are 
linked when they jointly appear in a significant way in the same comic 
book.  We only consider here comics published after Issue~1 of 
Fantastic Four (dated November 1961), which is understood as the point 
of departure of the Marvel Age of Comics.

Any study like this one must be based on a database, which puts the 
main restriction to its scope.  In this case, the database we have 
used is the \emph{Marvel Chronology Project} (MCP), which, according 
to its creator, R. Chappell \cite{MCP}, catalogs every canonical appearance by 
every significant Marvel character.  Thus, the ``significant 
characters'' represented by nodes in our network and the ``significant 
appearances'' that yield the links in it are, actually, nothing but 
those characters and appearances currently included in the MCP 
database.  Nevertheless, all in all, this database collects over 
$96\,000$ appearances by more than $6\,500$ characters in about 
$13\,000$ comic books, and thus yields quite a complete picture of the 
Marvel Universe.  Although the MCP database is not finished (it has a 
main gap, as it does not include comic books published between early 
1993 and mid 1994, as well as some other minor ones) we believe that 
this does not affect in a significant way the results obtained in our 
analysis.

It is necessary to clarify what we understood by a character when 
building up MU. On the one hand, it is quite common for the same 
person in the Marvel Universe to take different personalities.  As an 
example, recall Hank Pym, one of the original and most popular 
Avengers: it has been known, in different periods, as the Ant-Man, the 
Giant-Man, Goliath, YellowJacket, and has even appeared simply as the 
world's greatest biochemist Dr.\ Henry Pym in many books.  On the 
other hand, from time to time different characters may assume the same 
personality: for instance, besides Hank Pym, there have been at least 
two more Goliath's: Clint Barton (who changed from Hawkeye to Goliath, 
before returning back to Hawkeye) and Erik Josten (who was Power Man 
before becoming the third Goliath, and after that he took the name of 
Atlas, being actually the second character with this nickname).  In 
fact, these problems with the identification of nodes are not specific 
to MU, but they are shared by all collaboration networks: different 
authors can appear under the same name in a scientific collaboration 
network, and an actress could use a nickname during her period as 
prodigy child, then use her maiden name after adolescence, and then 
take her husbands' name after every wedding, coming back to her maiden 
name in every period between marriages.
Fortunately, and contrary to scientific databases or the Internet 
Movie Database, the MCP database takes care of most vicissitudes 
concerning name changes.  We decided then to assign a node to every 
``person'' (or, more in accordance with the nature of some characters, 
``entity''), independently of the nickname or personality under which 
it appears in each comic book.  In this way we have obtained $6\,486$ 
nodes, appearing in $12\,942$ comic books.

\section{Analysis of the network}

From the data contained in the MCP database, we have built up a 
bipartite graph (also known as \emph{mode 2 graph}), with nodes corresponding 
to either Marvel characters or comic books, and edges from every 
character to all the books where it has appeared.  We have extracted 
then from this bipartite graph the MU network, as its projection on 
its set of characters, and we have used PAJEK, a program for large 
network analysis \cite{Pajek}, to compute most of the key values in 
our study of MU. In this section we discuss in some detail the results 
we have obtained, which are numerically summarized in Tables~1 and 2.

\begin{table}[htb]
	\centering
	\caption{Basic data on appearances of characters in comic books.}
\begin{tabular}{ll}
Number of characters: & $6\,486$\\
Number of books: & $12\,942$ \\
Mean books per character: & $14.9$\\
Mean characters per book: & $7.47$ \\
Distribution of characters per book: & $P_{b}(k)\sim k^{-3.12}$\\
Distribution of books per character: & $P_{c}(k)\sim k^{-0.66} 
10^{-k/1895}$\\
\end{tabular}
\end{table}

\subsection{The bipartite graph}
The bipartite graph summarizing the MCP database contains 
$6\,486$ nodes corresponding to characters and $12\,942$ nodes 
corresponding to comic books, and $96\,662$ edges going from the characters 
to the books where they appear.

A Marvel character appears typically in about $14.9$ comic books.  The 
number of appearances spans from 1 to $1\,625$: this greatest value 
corresponds to Spider-Man.  The average number of characters per comic 
book is $7.47$ with a range spanning from 1 to 111: this last value is 
achieved by Issue~1 of \textsl{Contest of Champions} (1982), where the 
Grandmaster and the Unknown took every superhero in the planet and 
selected two teams to battle it out.

\begin{figure}[htb]
		\centering
\includegraphics[scale=0.75]{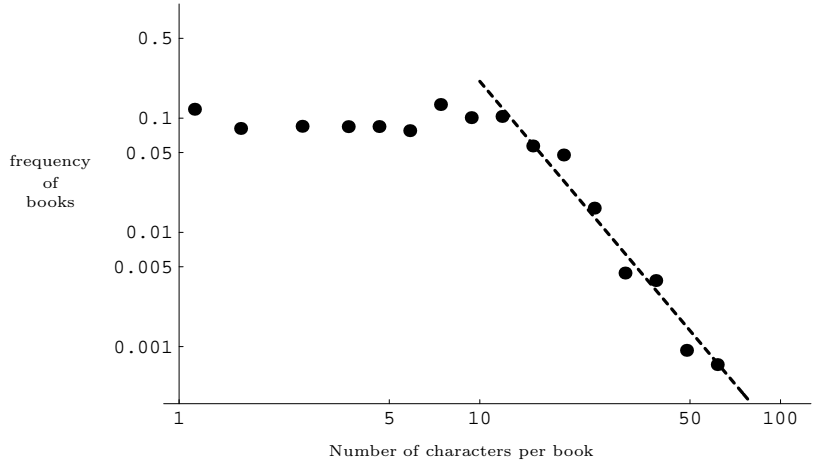}

\begin{picture}(0,10)
\put(0,10){\makebox(0,0)[t]{\tiny Number of characters per book}}
\put(-140,110){\makebox(0,0)[r]{\tiny {\shortstack{frequency\\ of\\ 
books}}}}
\end{picture}
		\caption{Distribution of characters per comic books in the 
		bipartite graph.  The horizontal axis corresponds to the 
		number of characters that appear in a comic book, while the 
		vertical axis represent the frequency of books with those many 
		characters.  Note that the scales on both axis are 
		logarithmic.  The dashed line shows the tail probability 
		distribution $P_{b}(k)\sim k^{-3.12}$.}
		\label{fig:revm2}
\end{figure}

We shall denote by $P_{b}(k)$ the distribution of ingoing edges, and 
by $P_{c}(k)$ the distribution of outgoing edges in this bipartite 
graph.  That is, $P_{b}(k)$ represents the probability that a comic 
book has $k$ characters appearing in it, and $P_{c}(k)$ represents the 
probability that a character appears in $k$ comic books. To obtain the 
best fit of these distributions we have logarithmically binned the 
data and performed a linear regression of $\log(P(r))$ on $\log(r)$.  
We have found that $P_{b}(k)$ follows the power-law tail %
$$
P_{b}(k)\sim k^{-3.1228}.
$$
The resulting histogram, together with the tail distribution is shown in 
Figure~\ref{fig:revm2}.  The distribution of $P_{b}$ is similar to 
what can be found in real-life networks and is a new example of the 
ubiquity of Zipf's law.

\begin{figure}[htb]
		\centering
\includegraphics[scale=0.75]{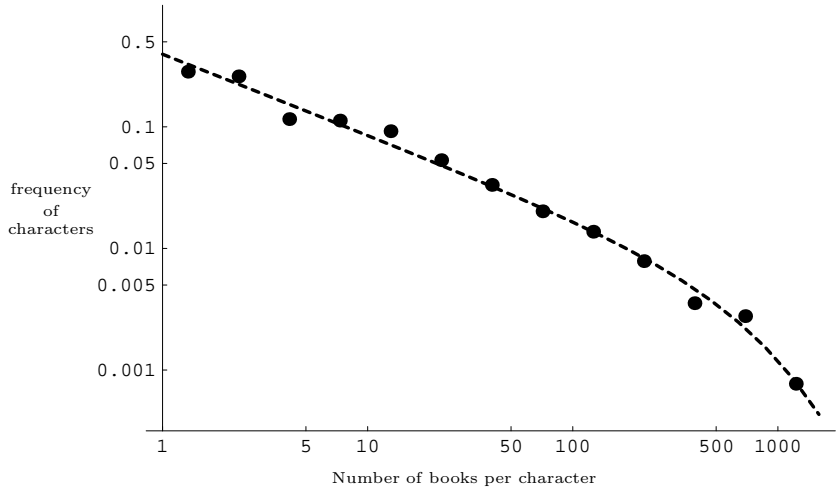}

\begin{picture}(0,10)
\put(0,10){\makebox(0,0)[t]{\tiny Number of books per character}}
\put(-140,110){\makebox(0,0)[r]{\tiny {\shortstack{frequency\\ of\\ 
characters}}}}
\end{picture}
		\caption{Distribution of books per character in the bipartite 
		graph.  The horizontal axis corresponds to the number of 
		comic books in which a character appears, while the vertical 
		axis represents the frequency of characters that appear in 
		those many books.  Note that the scales on both axis are 
		logarithmic.  The dashed line shows the probability 
		distribution $P_{c}(k)\sim k^{-0.66} 10^{-k/1895}$.}
		\label{fig:actm2}
\end{figure}

On the other hand the best fit for the distribution of $P_{c}(k)$ is 
different of what is normally found in bipartite graphs 
associated to collaboration networks.  The best fitting distribution 
we found is 
$$
P_{c}(k)\sim k^{-0.6644} 10^{-k/1895}.
$$
The exponent of only 0.66 is 
much smaller than other values published for similar networks, that 
usually ranges from 2 to 3.  Also, the presence of a cutoff has been 
seldom reported in the literature. It is also of note that the fitting is 
not only of the tail, but of all the histogram, with a high 
correlation of 0.992.  The histogram  together 
with the distribution found is shown in Figure~\ref{fig:actm2}.

These distributions will be the starting point to create a null random 
model against which to compare the characteristics of the Marvel Universe 
network. This model will be described in the next section.

\subsection{The null random model}

To gain some perspective on the results obtained from the MU network, 
we compare them to a null random model.  
A reasonable random model would seem to be one with its same set of 
nodes and whose links have been generated by simply tossing a 
(possibly charged) coin: each link exists, independently of the other 
ones, with a fixed probability $p$.  We shall call this a \emph{random 
network}.  Adjusting $p$, we can create a random network with as many 
nodes as our network and with expected number of links equal to the 
number of links in our network.  This null model is quite popular and  
has been often times used.

Recently, Newman et al.\ \cite{NWS}
have stated  that 
given that collaboration networks are created from bipartite graphs, a 
better null random model from which our expectations about network 
structure should be measured is obtained by projecting random 
bipartite graphs with predetermined distributions of ingoing and 
outgoing edges.  We have followed this approach in this paper.  More 
specifically, the null random model \emph{MU-R} we are going to 
compare the MU network to is obtained in the following way.  We start 
from a random bipartite graph, which we shall call a \emph{MU-BR} 
graph, with $6\,486$ nodes-characters and $12\,942$ nodes-books, and 
whose edges have been randomly created following exactly the same 
distributions $P_{c}(k)$ and $P_{b}(k)$ of outgoing and ingoing edges 
as those of the bipartite graph obtained from the MCP database in the 
previous subsection.  Then, a \emph{MU-R} graph is the projection of 
this random bipartite graph on its set of nodes-characters: i.e., 
its nodes correspond to characters and its links represent to be 
connected to the same book in a \emph{MU-BR} graph.  The theoretical 
data corresponding to this random model have been computed through the 
formulas given by Newman et al.\ in \textsl{loc.\ cit.}

\subsection{Basic data}
Our MU network has $N_{MU}=6\,486$ nodes (characters) and 
$M_{MU}=168\,267$ links, i.e. pairs of characters that have collaborated in 
some comic book.  We would like to mention that the actual number of 
collaborations is $569\,770$, but this value counts \emph{all} 
collaborations in the Marvel Universe history, and while there are 
$91\, 040$ pairs of characters that have only met once, other pairs 
have met quite often: for instance, every pair of members of the 
Fantastic Four has jointly appeared in around 700 comic books (more 
specifically, this range of collaborations of the members of the 
Fantastic Four runs between 668 joint appearances of the Thing and the 
Invisible Woman to 744 joint appearances of the Thing and the Human 
Torch).

The number of characters that have jointly appeared with a given 
character in some comic book is given by the \emph{degree} of this 
character in the network.  The average value for this degree in the MU 
collaboration network is 
$$
\frac{2M_{MU}}{N_{MU}}=51.88,
$$ 
i.e., a Marvel character has collaborated, on average, with 52 other 
characters.  The range of this number of collaborators runs from 0 
(the MCP database contains characters that have appeared in comic books where 
no other character is reported to appear) to $1\,933$, the number of 
partners of Captain America.

Even in such a basic quantity as the number of links, or, 
equivalently, the average degree of a node, we find a big difference 
between the values obtained in the MU network and in its null random 
model MU-R. Indeed, according to Newman et al.\ \cite[Eq.~(72)]{NWS}, in the MU-R 
graph we would expect all $569\,770$ collaborations to form 
\textsl{different} links, which is about $3.4$ times the actual number 
of links in the MU network.  As a consequence, the average degree in 
MU-R is the average degree in MU multiplied by this same factor, and 
would therefore become $175.69$: should the MU collaboration network 
(or, rather, the bipartite graph representing character appearances in 
books) have been created in a purely random way, a Marvel character 
would have collaborated on average with more than 175 other 
characters.

It is shown \cite[\S V.A]{NWS} that in the Hollywood graph and in 
several scientific collaboration networks the actual average degree is 
consistently smaller than the theoretical average degree of the 
corresponding random model, but not by such a large factor as the 
one found here.  This indicates that Marvel characters are made to 
collaborate  
repeatedly with the same characters, which reduces their total number 
of co-partners well below the expected number in the random model, and that 
they collaborate quite more often with the same people than real movie 
actors or scientists do.  This probably should be a hint of the 
artificiality of the Marvel Universe.

\begin{table}[htb]
	\centering
	\caption{Summary of results of the analysis of the MU network.}
\begin{tabular}{ll}
Mean partners per character: & $51.88$\\
Size of giant component: & $6\,449$ characters ($99.42$\%) \\
Mean distance: & $2.63$\\
Maximum distance: & 5\\
Clustering coefficient: & $0.012$\\
Distribution of partners:  & $P(k) \sim k^{-0.72} 10^{-k/2167}$  \\
\end{tabular}
\end{table}

\subsection{The giant component}
Two nodes in a network are said to be \emph{connected} when there is 
at least one path in the network, made of consecutive links, that 
connects them.  In a collaboration network, this means that two nodes 
are connected when they can be linked through a path of intermediate 
collaborators, or rather, in our case, co-partners.  As mentioned 
before, finding such a path, and more specifically the shortest one, 
between any actor or actress and Kevin Bacon in the Hollywood network, 
is the goal of the Kevin Bacon game.

In general, two nodes in a collaboration network need not be 
connected.  But, in all large enough, sensible real-life networks, 
almost every node is connected to almost every other node.  More 
specifically, large collaboration networks (and other large social 
networks) usually contain a very large subset of nodes ---around 80\% 
to 90\% of all nodes--- that are connected to each other: when this
happens, this large subset of nodes with their corresponding links is 
called the \emph{giant component} of the network. Also, 
Newman et al.\  \cite{NWS} show that in random collaboration networks 
giant components do also occur, provided the corresponding 
random bipartite graphs have enough edges.

MU contains a giant component of $6\,449$ nodes, which cover $99.42$\% 
of the characters in it.  Let us also mention that the
largest group of connected characters in the MU network outside this 
giant components has only 9 members.

\subsection{Separation}
The \emph{distance} between two connected nodes in a network is 
defined as the length (the number of links) of the shortest path 
connecting them, i.e., the least number of links we have to traverse 
in order to move from one node to the other within the network.  
Notice that the number of links in a path is equal to the number of 
intermediate nodes plus one, and thus we could also say that the 
distance between two connected nodes is the least number of 
intermediate nodes visited by a path connecting them plus one.  For 
instance, the Kevin Bacon game asks for the distance of any actor or 
actress to Kevin Bacon in the Hollywood network, as the least number 
of intermediate co-partners plus one linking that actor or actress to 
Kevin Bacon. And it is popular among mathematicians to compute 
\emph{Erd\"os numbers}, that is his/her distance to P. Erd\"os in the 
mathematicians' collaboration network.

We have calculated all distances between all pairs of connected nodes 
in MU. The greatest distance between two connected nodes, called the 
\emph{diameter} of MU in the usual network-theoretical terminology, is 
5.  It implies that there is always a chain of at most 4 collaborators 
connecting any two connectable characters in the Marvel Universe.

We have also computed the mean of all distances in the network, which 
provides the average separation of two characters in it.  The value of 
this average separation is $2.63$.  Thus, on average, any pair of 
characters in the MU network can be connected through a path of at 
most two consecutive partners.  This is larger than the expected value 
in the MU-R network, which is $1.45$.  Again, the reason is that only 
a third of the links in MU-R do appear in MU. Nevertheless, the 
values of both the diameter and the average separation in the MU 
network are significantly smaller than the values of real-life 
networks reported so far.

Finally, we have computed the \emph{center} of the giant component, 
the character that minimizes the sum of the distances from it to all 
other nodes in the component.  It turns out to be Captain America, who 
is, on average, at distance $1.70$ to every other character.

\subsection{Clustering}
\label{subsec-clust}
In most social networks, two nodes that are linked to a third one have 
a higher probability to be linked between them: two acquaintances of a 
given person probably know each other.  This effect is measured using 
the \emph{clustering coefficient}, that is defined as follows.  Given 
a node $v$ in a network, let $k_{v}$ be its degree, i.e., the number 
of neighbors of $v$, and let $N_{v}$ be the number of links between 
these $k_{v}$ neighbors of $v$.  If all these nodes were linked to 
each other, then $N_{v}$ would be equal to the number of unordered 
pairs of nodes belonging to this set of $k_{v}$ neighbors, i.e., to 
$k_{v}(k_{v}-1)/2$.  The clustering coefficient $C_{v}$ of node $v$ 
rates the difference between the actual value $N_{v}$ and this 
greatest value by taking their quotient
$$
C_{v}=\frac{2N_{v}}{k_{v}(k_{v}-1)}.
$$
Thus, this coefficient $C_{v}$ measures the fraction of neighbors of 
node $v$ that are linked. Notice that $0\leq C_{v}\leq 1$.
The \emph{clustering coefficient} $C$ of a network is then defined as 
the mean value of the clustering coefficients of all its nodes.  It 
represents the probability that two neighbors of an arbitrary node are 
linked.

All collaboration networks studied so far, and in general most social 
networks, have large clustering coefficients.  For instance, the 
clustering coefficient of the Hollywood network is 
$0.199$,\footnote{This figure is its last value, published by 
Newman et al.\ \cite{NWS}, and is quite different from the figure $0.79$ previously 
published by Watts \cite{Watts} when the network was quite smaller.} showing 
that two actors that have collaborated (possibly in different films) 
with a third actor, have greater probability of being partners in a 
movie than two arbitrary, randomly chosen, actors.  A similar effect 
appears in scientific collaboration networks: except for MEDLINE, all 
other scientific collaboration networks studied so far have their 
clustering coefficients between, roughly, $0.3$ and $0.8$, which tells 
us that a large fraction of the collaborators of a scientist 
collaborate with each other.  This large clustering, 
together with a low value of the average distance between connected 
nodes, is taken as the definition of \emph{small-world} networks 
\cite{Watts}.

Actually, the word ``large'' means large compared to the expected 
value of the clustering coefficient in a null random model.  
Depending on the choice of the null random model the results 
differ. It is worthwhile to dedicate some time to discuss the 
differences as this will shed some light on the nature of the Marvel 
Universe and how it differs from real-life collaboration networks.

In a random network with $n$ nodes and $m$ links, it can be proved 
that the expected value of the clustering coefficient is nothing but 
the probability $p$ that two randomly selected nodes are connected; 
in other words,
$$
C_{\mathrm{random}}=\frac{2m}{n(n-1)}.
$$
Measured values in collaboration networks are usually ``large'' in the 
sense that they are a few orders of magnitude larger than the 
predicted value of a random network.  For instance, the clustering 
coefficient of the MEDLINE network is $0.066$, which seems small, but 
it becomes very large when compared with the value $0.0000042$ of the 
clustering coefficient of a random network with the same number of 
nodes and links.

Against what happens with real-life social networks, it turns out that 
the clustering coefficient of MU is small, even in the last sense.  
Its value is $C_{\mathrm{Marvel}}=0.012$, while the clustering 
coefficient of a random network with $6\,486$ nodes and $168\,267$ 
links is $C_{\mathrm{random}}=0.008$.  Thus, roughly $C_{\mathrm{Marvel}}$ 
is $1.5\times C_{\mathrm{random}}$, and not several orders of 
magnitude larger.

This result separates the Marvel Universe from all other, real-life, 
collaboration networks.  But if we use the MU-R as null random network 
to compare the clustering coefficient of the MU network the analysis 
changes quite drastically.  The expected value of the clustering 
coefficient of the null random model MU-R using the formula given by 
Newman et al.\ \cite{NWS}, is $C_{\mathrm{MU-R}}=0.0066$.  Thus, the 
measured clustering coefficient is about double the one predicted by 
MU-R
$$
C_{\mathrm{Marvel}}\approx 2\times C_{\mathrm{MU-R}},
$$
and this agrees with what is observed in real-life networks, as shown 
by Newman et al.\ \cite[Table I]{NWS}: the clustering coefficient of the Hollywood 
network and the MEDLINE and Los Alamos e-Print Archive collaboration 
networks are between twice and $2.3$ times the expected clustering 
coefficient of the corresponding null random model.  So, in this sense, 
the tendency to clustering in the Marvel Universe is similar to that 
of real-life collaboration networks.

Our analysis shows that the Marvel Universe behaves ``realistically'' 
when compared to MU-R, but not when compared to a random network.  
Real-life collaboration networks have as clustering coefficient 
roughly twice the one of their null random model, and the latter turns 
out to be highly clustered.  The clustering coefficient of the MU 
network is also roughly twice the one of its null random model, but 
this null random model is not highly clustered, having a clustering 
coefficient only three times that of a random network with the same 
number of nodes and links.  We believe that, as we already argued in 
connection with the average degree, this is a hint of the 
artificiality of the bipartite graph which projects into MU. It seems 
that Marvel writers have not assigned characters to books in the same 
way as natural interactions would have done it, with the global effect 
that the combination of the distributions $P_{c}(k)$ and $P_{b}(k)$ is 
very different from what would be found in real-life networks, 
yielding non-clustered graphs.  But, once we have these distributions, 
the Marvel Universe behaves realistically and is different in a 
significant way from a random network.

\subsection{Distribution of the number of partners}

An interesting statistical datum that can be used to distinguish 
random networks from non-random networks is the distribution $P(k)$ of 
degrees in the network.  For every positive integer $k$, let $P(k)$ 
denote the fraction of nodes in a given network that have degree $k$.  
In a random network with $n$ nodes and $m$ links, the expected value 
for $P(k)$ follows a binomial distribution
$$
P(k)={n-1\choose k} p^k(1-p)^{n-1-k}
$$
where
$$
p = \frac{2m}{n(n-1)}.
$$

Against these values, it has been observed \cite{ASBS-00,Barab-sf} 
that in all collaboration networks considered so far the distribution 
$P(k)$ has a tail that follows either a power law
$$
P(k)\sim k^{-\tau}
$$
for some constant, positive exponent $\tau$, or  a power 
law form with an exponential cutoff
$$
P(k)\sim k^{-\tau} 10^{-k/c}
$$
where $\tau$ and $c$ are two positive constants, and $c$ is large.  
While the power law component of such a distribution allows the 
existence of a non-negligible number of nodes with high degree, the 
cutoff prevents the existence of nodes with very high degree.  In most 
collaboration networks, this cutoff is explained because the 
collaboration under consideration can only take place in a finite 
amount of time (for instance, the professional lifetime of an actor or 
a scientist), which makes unplausible the existence of nodes with a 
number of collaborators greater than some reasonable upper bound.

\begin{figure}[htb]
		\centering
\includegraphics[scale=0.75]{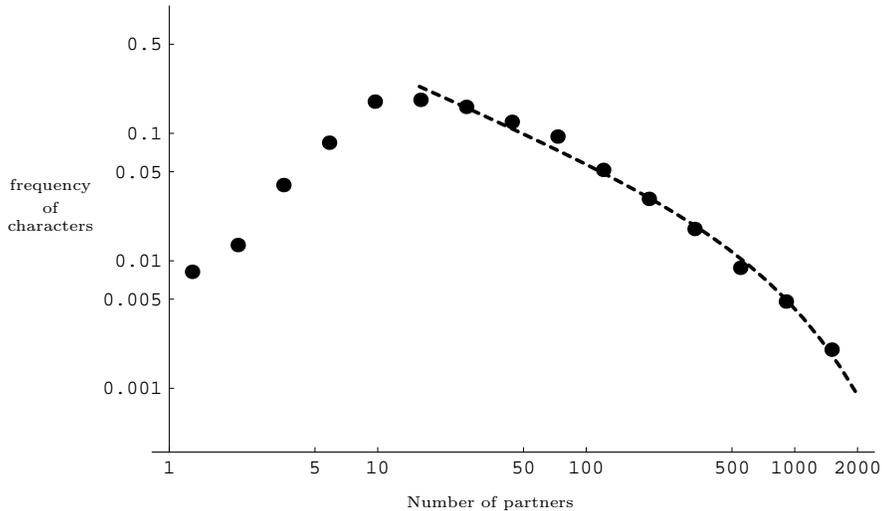}

\begin{picture}(0,10)
\put(0,10){\makebox(0,0)[t]{\tiny Number of partners}}
\put(-150,120){\makebox(0,0)[r]{\tiny {\shortstack{frequency\\ of\\ 
characters}}}}
\end{picture}
		\caption{Degree distribution in the MU network. The horizontal 
		axis corresponds to the number of relations (or links) of a character, while the 
		vertical axis represent the frequency of characters with those many 
		relations. Note that the scales on both axis are logarithmic. The 
		dashed line shows the tail probability distribution 
		$P(k) \sim k^{-0.7158} 10^{-k/2167}$.}
		\label{fig:actm1}
\end{figure}

When we tested how the data fit a power law distribution with an 
exponential cutoff, we obtained that the best fitting tail was $P(k) 
\sim k^{-0.7158} 10^{-k/2167}$.  Thus, the degree distribution of the 
MU network has a power law tail with cutoff as can be seen in 
Figure~\ref{fig:actm1}.

A value of $\tau$ smaller than 2 means that the average properties of 
the network are dominated by the few actors with a large number of 
collaborators.  This happens in many real-life networks \cite{N1}.  
The value of $\tau$ much smaller than 2 shows that the weight of 
Captain America, Spider-Man, and other major super-heroes is much 
larger than what happens in scientific or movie actor collaborator 
networks.  This should be expected: there are no super-heroes in real 
life.

\section{Conclusions}

Real-life collaboration networks of very different origins, sizes and 
styles present common basic features: they are scale-free, small 
worlds.  We have tried to ascertain if these characteristics depend on 
some profound social relationship or are there by chance.  We have 
studied the Marvel Universe, which is a collaboration network that is 
artificial and has been created with no special intention during the 
past 40 years by a team of comic book writers. 

Although to some extent the Marvel Universe tries to mimic human 
relations, and in particular it is completely different from a random 
network, we have shown that it cannot completely hide its artificial 
origins.  As in real-life collaboration and, in general, social 
networks, its nodes are on average at a short distance of each other, 
and the distribution of collaborators shows a clear power-law tail 
with cutoff.  But its clustering coefficient is quite smaller than 
what's usual in real-life collaboration networks.
 
We have compared the Marvel Universe network with a null random model 
obtained as the projection of a bipartite random graph with the same 
number of character and comic-book nodes and the same distribution of 
ingoing and outgoing edges as the actual bipartite graph of 
appearances of characters in books which the Marvel Universe is built 
upon.  From this comparison we deduce that the artificiality of the 
Marvel Universe network lies mainly on the distributions of edges in 
the bipartite graph which yields it, because the relationship between 
the Marvel Universe network's data and those of its null random model 
is similar to that of real-life collaboration networks' data and their 
corresponding null random models.
 
From here we conclude that in the construction of real collaboration 
networks there are two unknown, profound different principles in play.  
On the one hand, the degree distributions of the bipartite graph which 
they are based upon are not arbitrary.  On the other hand, the final 
structure of any actual collaboration network, be it real-life or 
artificial, differs from its null random collaboration network model 
roughly in the same way, and thus there is probably a common mechanism 
that produces them.  Further study is needed to find what these 
principles may be.

We believe that continuing to study the Marvel Universe may contribute 
to this search.  Namely, the study of its evolution and its comparison 
with that of real-life collaboration networks should shed some light 
on the basis of the aforementioned mechanism and on where the 
artificiality of this network lies.  Fortunately, the MCP database, in 
the words of its creator R. Chapell, attempts to ``not only catalog 
every canonical appearance by every significant Marvel character, but 
to place those appearances in proper chronological order,'' and thus 
it contains enough information to allow such study.  We hope to report 
on it in the future.

\bibliography{smallworlds}
\bibliographystyle{plain}

\end{document}